\documentclass[footinbib,pre,aps,showpacs,bibnotes,twocolumn]{revtex4}

\usepackage{inputenc}
\inputencoding{ansinew}
\usepackage[T1]{fontenc}
\usepackage{graphicx}
\usepackage[Gray,squaren]{SIunits}
\usepackage{amsmath}
\usepackage{amssymb}
\usepackage{bm}
\usepackage{pstricks}
\usepackage{nicefrac}

\newcommand*{\E}[1]{\cdot 10^{#1}}
\renewcommand{\Re}{\mathrm{Re}}
\newcommand{\omegae}{\omega_{\text{e}}}
\newcommand{\nue}{\nu_{\text{e}}}

\begin{document}
\title{Frequency Dependent Specific Heat from Thermal Effusion in
  Spherical Geometry.}  
\author{Bo Jakobsen}
\author{Niels Boye Olsen} 
\author{Tage Christensen}
\email{tec@ruc.dk}
\affiliation{DNRF centre ``Glass and Time,'' IMFUFA,
  Department of Sciences, Roskilde University, Postbox 260, DK-4000
  Roskilde, Denmark} 

\date{\today}

\begin{abstract}
  We present a novel method of measuring the frequency dependent
  specific heat at the glass transition applied to
  5-polyphenyl-4-ether. The method employs thermal waves effusing
  radially out from the surface of a spherical thermistor that acts as
  both a heat generator and thermometer. It is a merit of the method
  compared to planar effusion methods that the influence of the
  mechanical boundary conditions are analytically known. This implies that it is
  the longitudinal rather than the isobaric specific heat that is
  measured. As another merit the thermal conductivity and specific
  heat can be found independently. The method has highest sensitivity
  at a frequency where the thermal diffusion length is comparable to
  the radius of the heat generator. This limits in practise the
  frequency range to $2$--$3$ decades. An account of the
  $3\omega$-technique used including higher order terms in the
  temperature dependency of the thermistor and in the power generated
  is furthermore given.
\end{abstract}


\maketitle

\section{Introduction}
\label{Introduction}
A hallmark of the glass transition of supercooled liquids is the time
dependence of physical properties --- also called relaxation. This time
dependence is typically seen on the timescale of the so-called Maxwell
relaxation time $\tau_M$. $\tau_M$ is defined as
\begin{equation}
 \tau_M=\frac{\eta_0}{G_\infty}
\end{equation}
where $\eta_0$ is the low frequency limiting (DC) shear viscosity and
$G_\infty$ the high frequency limiting elastic shear modulus. The relaxation
is often most easily studied in the frequency domain. At high
frequencies the liquid shows solid-like elastic behaviour whereas at
low frequencies it shows liquid-like viscous behaviour. In general the
viscoelastic stress response to oscillatoric shear strain is
described by the complex dynamic shear modulus $G$ \cite{har76}. The
structural relaxation is also observed in the specific heat. This has
long ago been studied in the time domain as enthalpy relaxation
\cite{Dav53}, but in 1985 frequency domain studies or AC-calorimetry
\cite{bir85, Chr85} were introduced in the research field of
supercooled liquids.  Common to these methods is that they excite
thermal oscillations in the sample at a cyclic frequency $\omega$. For
a heat diffusivity, $D$, this frequency is associated with a
characteristic heat diffusion length,
\begin{equation}
 l_D=\sqrt{\frac{D}{i\omega}},
\end{equation}
the magnitude of which gives the range of the excited thermal waves.
When comparing $|l_D|$ to a given sample size, $L$, one can classify
AC-calorimetry experiments as to whether the sample is thermally thin
($L\ll |l_D|$) or thermally thick ($L\gg|l_D|$). A typical value of
$D$ is $0.1\text{mm}^2/\text{s}$ giving $|l_D|\approx 0.1\text{mm}$
for a frequency of $\nu=\omega/(2 \pi)=1\text{Hz}$. This means that
thermally thin methods are restricted to rather low frequencies
\cite{Chr85,Christensen1998} or very thin samples
\cite{Huth07,Christensen1998}. In these methods the temperature field
is homogeneous throughout the sample and the frequency dependent
specific heat is derived directly. The AC-technique allows for easy
corrections of the influence of heat leaks in a non-adiabatic
configuration. AC-calorimetry on thermally thick samples on the other
hand has the advantage of covering a large frequency span
\cite{bir85,bir86}. In this case the temperature field is inhomogenous
and what is really measured is the effusivity, $e=\sqrt{\lambda c}$,
from which the specific heat is derived. The effusivity expresses the
ability of the liquid to take up heat from parts of its surface and
transport it away (thermal effusion). This ability is dependent on two
constitutive properties, the heat conductivity, $\lambda$, and the
specific heat (pr. volume), $c$. The specific heat can only be found to
within a proportionality constant in effusion experiments from an
(infinite) plane. However taking boundary effects into account for
finite size plane heaters \cite{Moon96, Birge97} one may in principle
find $\lambda$ and $c$ independently.  The intermediate situation
where the heat diffusion length is comparable to the sample thickness
has been exploited in the so-called two-channel ac calorimeter
\cite{Min01,Min03}. This fine method aims directly at an independent
determination of $\lambda$ and $c$.

The specific heat entering the effusivity is widely taken as the
isobaric specific heat, $c_p$. However recently it has been shown
\cite{chr07,chr08,chr08b} that in planar and spherical geometries it
is rather the longitudinal specific heat, $c_l$, that enters. When $M_S$
and $ M_T$ are the adiabatic and isothermal longitudinal moduli
respectively, $c_l$ is related to the isochoric specific heat, $c_V$,
by
\begin{equation}\label{eq:cldeff}
 c_l=\frac{M_S}{M_T}c_v.
\end{equation}
In contrast, when $K_S$ and $ K_T$ are the adiabatic and isothermal
bulk moduli respectively, $c_p$ is given by
\begin{equation}\label{eq:cpdeff}
 c_p=\frac{K_S}{K_T}c_v.
\end{equation}
Now $M_S=K_S+4/3\,G$, and $M_T=K_T+4/3\,G$. This means that $c_l$ will
differ from $c_p$ in the relaxation regime where the shear modulus,
$G$, dynamically is different from zero, and $K_s$ and $K_T$ differ
significantly.

In this work we study the frequency dependent specific heat from
thermal effusion in a spherical geometry.  Specific heat measurements
in a spherical geometry have earlier been conducted in the thermal
thin limit \cite{Christensen1998}, but in this paper we explore the
thermal thick limit.
 
The justification is threefold 1) We want to realize an experimental
geometry in which the thermomechanical influence on the thermal
effusion can be analytically calculated \cite{chr08b}.  2) The
spherical container of the liquid is a piezoelectric spherical shell
by which it is possible to measure the adiabatic bulk modulus
\cite{Chr94} on the very same sample under identical conditions.  3)
The heat conductivity and specific heat can be found independently, as
the introduction of a finite length scale allows for separation of the
two.

We also in this paper give an account of the $3\omega$ detection
technique of our variant.

\section{Linear Response Theory Applied to Thermal Experiments}
\label{sec:LinResponseTheory}
The experimental method used in this study is an effusion method. In
such methods a harmonic varying heat current, $\Re(Pe^{i\omega t})$,
with angular frequency $\omega$ and complex amplitude $P$, is produced
at a surface in contact with the liquid; this heat then ``effuses''
into the liquid.  At the surface a corresponding harmonic temperature
oscillation, $\Re(\delta Te^{i\omega t})$, with complex amplitude
$\delta T$, is created.  The current and temperature can be thought of
as stimuli and response (but which is which can be interchanged at
will), and the two are linearly related if the amplitudes, $|P|$ and
$|\delta T|$, are small enough. Hence it is convenient to introduce
the complex frequency-dependent thermal impedance of the liquid
\begin{eqnarray}
  \label{eq:Zthermal}
  Z_\text{liq}=\frac{\delta T}{P},
\end{eqnarray}
this impedance depends on the properties of the liquid.

The terminology thermal \textit{impedance} derives naturally from the
close analogy between the flow of thermal heat and the flow of
electricity \cite{car59}. Temperature corresponds to electric
potential and heat current to electrical current. The analogy also
applies to the heat conductivity which corresponds to the specific
electrical conductivity, and the specific heat which corresponds to
electrical capacity.

As an example the thermal impedance of a thermally thin sample is
given as
\begin{eqnarray}
  Z_{\text{liq,thin}}=\frac{1}{i\omega c V}  
\end{eqnarray}
where $V$ is the volume of the sample. Again an equation completely
equivalent to the electrical correspondence between capacitance and
impedance.

In the classical case of a planar heater in a thermal thick situation
the thermal impedance is found to be \cite{bir85}
\begin{eqnarray}
  \label{eq:Zplane}
  Z_\text{liq,planar}=\frac{1}{A\sqrt{i\omega c \lambda}}
\end{eqnarray}
where $A$ is the plate
area. The lateral dimension $W$ has to be large ($W \gg
|l_D(\omega)|$) for this formula to be exact and unfortunately
correction terms that take boundary effects into account decays only
slowly as ${\omega}^{-1/2}$ \cite{Birge97}.

\section{Thermal Effusion in Spherical Geometry}
\label{sec:ThrmEffSphGeo}
Thermal effusion is possible in a spherical geometry as well as in a
planar geometry. Imagine a liquid confined between the to radii, $r_1$ and
$r_2$. The imposed oscillating outward heat current, $\Re\left(P(r_1)
  e^{i\omega t}\right)$, at radius $r_1$ results in a temperature
oscillation, $\Re\left(\delta T(r_1)e^{i\omega t}\right)$, at the same
radius. The heat wave propagating out into the liquid gives rise to a
coupled strain wave due to thermal expansion. The heat diffusion is
thus dependent on the mechanical boundary conditions and mechanical
properties of the liquid. It is widely assumed that the transport of heat in a
continuum is described by the heat diffusion equation
\begin{equation}\label{eq:TESG1}
i\omega \delta T=D\nabla^2 \delta T.
\end{equation}
By the same token one might take the heat diffusion constant as
$D=\lambda/c_v$ if the surfaces at $r_1$ and $r_2$ are clamped and
$D=\lambda/c_p$ if one of the surfaces is free to move. This common
belief is not entirely correct. If shear modulus is nonvanishing
compared to bulk modulus the description of heat diffusion cannot be
decoupled to a bare heat diffusion equation
\cite{lan86,chr07,chr08,chr08b}. Such a situation arises dynamically
at frequencies where $\omega\approx \nicefrac{1}{\tau_M}$.  The fully
coupled thermomechanical equations have been discussed in detail, by
some of us, in planar geometry \cite{chr07} and in spherical geometry
\cite{chr08b}.  In general the solutions are complicated but in the
case of a spherical geometry where the outer radius $r_2$ is much
larger than the thermal diffusion length, $|l_D(\omega)|\ll r_2$ the
solution becomes simple.  In terms of the liquid thermal impedance,
$Z_\text{liq}$, at the inner radius, $r_1$, one has
\begin{equation}\label{eq:TESG2}
  Z_{\text{liq}}=\frac{\delta T(r_1)}{P(r_1)}=\frac{1}{4 \pi \lambda r_1 \left(1+ \sqrt{i\omega r_1^2 c_l/\lambda}\right)}\,,
\end{equation}
It is very interesting that in this thermally thick limit, the thermal
impedance is independent of the mechanical boundary conditions
\cite{chr08,chr08b}. I.e.\ in the two cases of clamped or free
boundary conditions mentioned before it is neither the isochoric nor
the isobaric specific heat that enters the diffusion constant and the
thermal impedance but the longitudinal specific heat. Since a closed
spherical surface has no boundary there are no correction terms like
the ones to Eq.\ (\ref{eq:Zplane}) discussed above for planar plate
effusion. Notice from Eq.\ (\ref{eq:TESG2}) that associated with a given
radius, $r_1$, is a characteristic heat diffusion time for the liquid
\begin{equation}
 \tau_l={r_1}^2 c_l/\lambda.
\end{equation}
$c_l$ and hence $\tau_l$ may in general be complex, but when 
$c_l$ is real, the imaginary part of $Z_{\text{liq}}$ peaks at
$\omega=1/\tau_l$.

\section{$3\omegae$-detection technique}
\label{sec:3-omega-detection}
The $3\omegae$ detection technique we use (where the subscript $e$
indicates that it is the electrical angular frequency not the thermal)
is different from other $3\omegae$ techniques (e.g.\ Ref.\
\cite{Bir87}) in two ways. Firstly, we include higher order terms in
the temperature dependence of the thermistor used and in the power oscillations produced. Secondly, we use
a direct measurement of the time dependent voltage over a voltage
divider instead of a lock-in amplifier.

\subsection{The principle.}
The $3\omegae$-detection technique makes it possible to measure the
temperature with the same electrical resistor which generates the heat
and thereby know the thermal current and the temperature at the same
surface. The principle is in short the following: Applying an
electrical current at cyclic frequency $\omegae$ through a resistor a
combined constant (zero frequency) and $2\omegae$ component Joule
heating is produced. This gives rise to a combined temperature
DC-offset and temperature oscillation at $2\omegae$, the size of which
depends on the thermal environment (the thermal impedance).  The
resistor is chosen to be temperature dependent, it is a so-called
\textit{thermistor}. The temperature oscillations therefore creates a
perturbation of the resistance at frequency $2\omegae$. The voltage
across the thermistor which by Ohm's law is the product of current and
resistance thus contains a $3\omegae$ (and an additional $1\omegae$)
component proportional to the temperature oscillations at $2\omegae$.
In practise one rather has a voltage source than a current source. In
order to get a $3\omegae$ signal the thermistor is therefore placed in
series with a temperature independent preresistor as detailed in the
following.

The result of the analysis is the frequency dependent thermal
impedance at a frequency of $2\omegae$. It is convenient to express
this thermal impedance as function of the frequency of the thermal
signal which we designate $\omega$ and notice that $\omega=2\omegae$.

We first outline the simplest theory in order to elucidate the
technique most clearly. However, to increase the signal to noise ratio
it is expedient to choose a thermistor with a large temperature
dependence and to choose temperature amplitudes as high as possible
although within the linear regime of the liquid thermal response. This
necessitates higher harmonics be taken into account.  This detailed
analysis is deferred to appendix \ref{app:ACHigher}.

\subsection{A detour on the complex notation.}

Complex numbers are known to be of great use in harmonic analysis of
linear systems. Since the heat current is quadratic in the imposed
electrical current and the whole idea of the $3\omegae$-detection
technique is to exploit the inherent nonlinearity introduced by the
temperature dependence of the thermistor the formalism becomes a
little more troubled than in pure linear applications. Thus we have to
drag around the complex conjugate terms in order to get product terms
right. 

Generally a sum of harmonic terms,
\begin{equation}
A=A_0+|A_1| \cos(\omegae t+\phi_1)+\ldots+|A_n| \cos(n\omegae t+\phi_n),
\end{equation}
with real amplitudes, $A_0$,$|A_k|$, and phases, $\phi_k$, can be
written as,
\begin{multline} 
  A=\frac{1}{2} \left(A_0 +A_0+|A_1| e^{i(\omegae t + \phi_1)}+|A_1|
  e^{-i(\omegae t + \phi_1)}+ \ldots \right.\\  \left. + |A_n| e^{i(n\omegae t +
    \phi_n)}+|A_n| e^{-i(n\omegae t + \phi_n)}\right). 
\end{multline}
By introducing the complex amplitudes $A_k=|A_k| e^{i\phi_k}$ and the
shorthand notation $E_k=e^{ik\omegae t}$ the sum becomes
\begin{equation}
A=\frac{1}{2}(A_0+A_1 E_1+\ldots+A_n E_n +c.c.),
\end{equation}
where $+ c.c.$ means that the complex conjugated of all the terms within
the parenthesis should be added, including the real constant term.
Note that $E_0=1$, $E_k E_l=E_{k+l}$, and $E_kE_l^*=E_{k-l}$.

Notice that any product of an $n$'th and $m$'th harmonic will produce
both an $n+m$'th and $n-m$'th harmonic term. Furthermore the DC-components
get doubled because there is no difference between $E_0$ and
$E_0^*$.

\subsection{Fundamental equations for the voltage divider}
\begin{figure}[t]
\begin{center}
\includegraphics[]{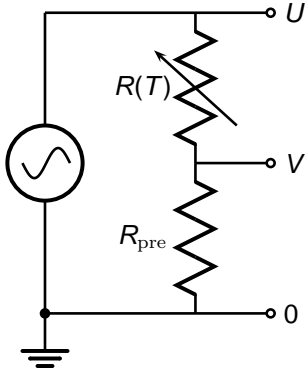}
\caption{Voltagedivider for the $3\omegae$-detection technique. The
  temperature-dependent resistor (the thermistor), $R(T)$, is in
  thermal contact with the liquid probing its thermal impedance at
  $2\omegae$. The preresistor, $R_\text{pre}$, is temperature-independent.}
\label{fig:vdiv}
\end{center}
\end{figure}

Consider the diagram of the electrical setup seen in Fig.\
\ref{fig:vdiv}.  $U(t)$ is the voltage of the source and $V(t)$ the
voltage across the temperature independent preresistor,
$R_\text{pre}$. $U(t)$ and $V(t)$ can be measured directly.  $R(T(t))$
is the temperature dependent resistance of the heater/sensor (we will
explicitly state the temperature/time dependencies when appropriate,
otherwise it is implicitly assumed).

The current $I(t)$ through $R(T(t))$ and $R_\text{pre}$ is given as
  \begin{eqnarray}  \label{eq:IofUbasic} \label{eq:IofVbasic}
    I(t)&=& \frac{U(t)}{R_\text{pre}+R(T(t))} = \frac{V(t)}{{R_\text{pre}}},
  \end{eqnarray}
and the power produced in the heater as
\begin{eqnarray}\label{eq:Pbasic}  
    P(t)=I(t)^2R(T(t))=\frac{R(T(t))}{\left( R_\text{pre}+R(T(t))\strut\right)^2}U(t)^2.
\end{eqnarray}

Eq.\ (\ref{eq:IofUbasic}) gives the voltage divider equation,
\begin{eqnarray}
  \label{eq:VoltageDiv}
  V(t)=\frac{R_\text{pre}}{R_\text{pre}+R(T(t))}U(t),
\end{eqnarray}
which in principle allows for determination of $T(t)$ from the
measured voltages.

\subsection{The AC-method to lowest order}
\label{sec:first-order-model}
The thermistor and the surrounding sample are placed in a cryostat
that defines an overall reference temperature, $T_\text{cryo}$. The
power produces a (real) temperature change $\Delta T=T-T_\text{cryo}$
at the heater/sensor depending on the thermal impedance, $Z$, (as
defined in section \ref{sec:LinResponseTheory}) the heater/sensor
looks into.  The temperature rise has a real DC-component, $T_0$, and
complex AC-components. It can in general be written as
\begin{equation} \label{DeltaTFourExp1}
 \Delta T=\frac{1}{2}(T_0+T_1 E_1+T_2 E_2+T_3 E_3+T_4 E_4 + ... +c.c.)
\end{equation}
the dominant terms will be the DC component, $T_0$, and the second
harmonic term, $T_2$ (see next section). 

To first order in the temperature variations the thermistor has a
resistance of
\begin{equation}  \label{eq:Rfirst}
 R=R_0(1+\alpha_1 \Delta T).
\end{equation} 
Here $R_0$ and $\alpha_1$ are considered constant at a given cryostate
temperature but they vary with $T_\text{cryo}$. In fact the thermistor,
which is based on a semiconducting material, has a resistance that over
a larger temperature range is very well described by
\begin{eqnarray}
  \label{eq:Rfull}
  R=R_{\infty}e^{T_a/T},
\end{eqnarray}
with $T_a$ the activation temperature and $R_{\infty}$ the infinity
temperature limiting resistance. $\alpha_1$ is related to $T_a$ by
\begin{equation}\label{eq:alpha1}
\alpha_1= \frac{1}{R(T_{\text{cryo}})}\left.\frac{d R}{d T}\right|_{T_\text{cryo}}= -\frac{T_a}{T_\text{cryo}^2}.
\end{equation}
Characteristic values of $R_\infty$ and $T_a$ are given in table
\ref{tab:ExpParm}.

\begin{table}
  \centering
  \caption{Experimental parameters. \textbf{Top:} Parameters
    characterizing the resistors in the setup, $R_\text{pre}$ is the
    resistance of the preresistor,  $R_{\infty}$ and $T_a$ the
    characteristics of the 
    thermistor according to Eq.\  (\ref{eq:Rfull})). 
\textbf{Middle:} Absolute size of the thermistor bead. Measured
values indicates the eccentricity of the bead. The optimize value is
the one use in the final treatment of the data, and is optimized to
give best correspondence between the room temperature DC specific
heat, and the literature value of this. 
\textbf{Bottom:} Applied amplitudes expressed as
    mean applied voltage, ($\langle |U_1|\rangle_{\nu}$), and maximum
    temperature amplitude, ($\max(|T_2|)_{\nu,T})$.} 
  \label{tab:ExpParm}
  \setlength{\tabcolsep}{0.7em}

   \begin{tabular}{ll}
                  \hline \hline
     $R_\text{pre}$   & $3884.3\ohm$   \\
     $R_{\infty}$& $0.1251\ohm$   \\
     $T_a$      & $2789.5\kelvin$ \\ \hline
     $r_{1,\text{Measured, long edge}}$ & $0.205\milli\meter$\\
     $r_{1,\text{Measured, short edge}}$ & $0.185\milli\meter$\\
     $r_{1,\text{Optimized}}$
 & $0.2073\milli\meter$ \\ \hline
     $\langle |U_1|\rangle_{\nu}$  &$4.909\volt$\\
     $ \max(|T_2|)_{\nu,T}$   &$2.253\kelvin$ \\ 
\hline \hline
  \end{tabular} \hspace{1cm}
\end{table}

Substituting Eq.\ (\ref{eq:Rfirst}) into Eq.\ (\ref{eq:VoltageDiv}) we
obtain
\begin{eqnarray}\label{eq:V1order}
  V=\frac{1}{1+A(1+\alpha_1 \Delta T)}U\approx\frac{1}{A+1}\left(1-a\Delta T\right)U 
\end{eqnarray}
where $A=R_0/R_\text{pre}$, $a=\frac{A\alpha_1}{1+A}$ and the second
expression is to first order in $\Delta T$ like  in Eq.\
(\ref{eq:Rfirst}).

In the simple case, which we first discuss, we assume that the function
generator can deliver a pure single harmonic voltage, 
\begin{equation}\label{UFourExp1}
 U(t)=\frac{1}{2}(U_1 E_1+ c.c).
\end{equation}

Eqs. (\ref{DeltaTFourExp1}) and (\ref{UFourExp1}) can now be
substituted into Eq.\ (\ref{eq:V1order}). 
\begin{multline}
  \label{eq:2}
  V=\frac{1}{A+1}\left(1-a\left(\frac{1}{2}(T_0+T_1 E_1+T_2 E_2+T_3
      E_3+  \right. \right. \\ \left. \left. \vphantom{\frac{1}{2}} T_4 E_4 + \ldots +c.c.)\right)\right)\frac{1}{2}(U_1 E_1+ U_1^* E_1^*).
\end{multline}

In principle the result contains all harmonics, but it turns out that
the interesting ones are the first and third harmonic. The complex
amplitudes are by inspection seen to be 
\begin{equation} \label{eq:V1first}
 V_1=\frac{1}{A+1}\left(U_1-aT_0U_1-\frac{1}{2}aT_2U_1^* \right)
\end{equation}
and
\begin{equation} \label{eq:V3firstFull}
 V_3=\frac{1}{A+1}\left(-\frac{1}{2}aT_2U_1 -\frac{1}{2}aT_4U_1^*
\right).
\end{equation}
Notice that even if $\Delta T$ is written as an infinite sum (as in
Eq.\ (\ref{DeltaTFourExp1})), $V_1$ and $V_3$ have only the included
terms. This is a consequence of assuming a perfect voltage source.

The $T_4$ term can be neglected, since it is proportional to a higher
order power term, $P_4$, which is small. This approximation leads to
the a use full relationship between the $3\omegae$-term and the
temperature amplitude at $2\omega$, $T_2$,
\begin{equation} \label{eq:V3first}
 V_3=-\frac{1}{2}\frac{a}{(A+1)}T_2U_1.
\end{equation}
Eq.\ (\ref{eq:V1first}) and (\ref{eq:V3first}) can be solved for $T_0$
and $T_2$ if prior knowledge on the temperature dependency of the
resistance of the thermistor exists, that is if $A$ and $a$ are known
at the relevant temperatures. The $T_2$ and $T_0$ component terms are
related to the power through the complex AC-thermal impedance, $Z_2$
(the impedance at $2\omegae$), and DC-thermal impedance, $Z_0$ as
\begin{equation} \label{eq:Z2def}
 T_0=Z_0 P_0\ \text{and}\  T_2=Z_2 P_2.
\end{equation}

The temperature dependency of the thermistor resistance can of course 
be found from calibration measurements. We have however found it to be
more efficient to use an iterative solution technique eliminating the
need for prior calibration. Eq.\ (\ref{eq:V1first}) and
(\ref{eq:V3first}) are rewritten in the following form
\begin{subequations} \label{eq:LowestOrderSolution}
 \begin{eqnarray}
  A&=& \frac{\left(U_1-aT_0U_1-\frac{1}{2}aT_2U_1^* \right)}{V_1}-1    \label{eq:Afirst}\\
   T_2&=&-2\frac{V_3(A+1)}{aU_1}.    \label{eq:T2first}
 \end{eqnarray}
\end{subequations}
It can be seen that the equation for $A$ consists of the $0$'th order
expression $A\approx \nicefrac{U_1}{V_1}-1$ which would be valid if
the temperature of the thermistor bead was the same as the cryostat
temperature. This is not the case as the thermistor generates self
heat, which is what the two additional terms correct for.

$A$ and $T_2$ can be found by iteration in the following way.  A first
approximation of $A$ is found by setting $T_0=0$ and $T_2=0$ in Eq.\
(\ref{eq:Afirst}). Based on this an initial estimate of $\alpha_1$ is
found from the temperature dependence of $A$. An estimate on $T_2$
(and hence $Z_2$) is afterwards found from Eq.\ (\ref{eq:T2first}). By
extrapolating $Z_2$ to zero frequency an estimate of $T_0$ is found
\footnote{The DC value of the thermal impedance, $Z_0$, can be
  estimated from the low frequency limit of the AC thermal impedance
  at $2\omegae$, $Z_2$. Our lowest frequency of measurement is
  $10^{-3}\hertz$ at which the thermal wavelength is still shorter
  than the sample size (we are still in the thermally thick
  limit). However the DC impedance does see the outer boundary, as it
  has ``infinite'' time to reach steady state. To compensate for this
  we add a small frequency, temperature and amplitude independent
  correction term to the limiting value of $Z_2$ when estimating
  $Z_0$. This correction terms value was chosen so that measurements
  with different amplitude, gives the same values for $A(T)$.}
(utilizing $T_0=Z_0 P_0$). The estimated values for $T_0,T_2$ and $a$
can then be inserted into Eq.\ (\ref{eq:Afirst}) giving a better
estimate for $A$. This iteration procedure can be repeated a number of
times and converges rapidly to a fixpoint for $A$ and $T_2$.

\subsection{Thermal-power terms}
\label{sec:power-terms}
If the 1.\ order approximation of the resistance of the thermistor,
Eq.\ (\ref{eq:Rfirst}), is substituted into the power equation, Eq.\
(\ref{eq:Pbasic}), the following expression is obtained to first order
in $\alpha_1\Delta T$
\begin{multline}
  \label{eq:P1order}
  P=\frac{1}{R_\text{pre}}\frac{A(1+\alpha_1 \Delta T)}{(1+A(1+\alpha_1
    \Delta T)\strut)^2}U^2 \approx \\
  \frac{1}{R_\text{pre}}\frac{A}{(1+A)^2}\left(1+\frac{1-A}{1+A}\alpha_1\Delta T \right)U^2.
\end{multline}

This shows that an input voltage which is a pure 1.\ harmonic leads to
dominant 0.\ and 2.\ harmonics in the power/temperature. Notice that if
the thermistor resistance were temperature independent, that is
$\alpha_1=0$, then $P_0$ equals $|P_2|$ exactly but in general they
are almost identical, $P_0\approx|P_2|$.

It can further be seen that even if the voltage generator is purely
single harmonic, the power can have a $4\omegae$ component from the
interplay between the $2\omegae$ temperature variation and the
$2\omegae$ variation of the squared voltage.  In principle
further higher harmonics are generated this way. It can however be
observed that if $\alpha_1\Delta T$ is small, or if $A\approx 1$ which 
can be chosen in the setup, the higher harmonics will be small (and decreasing
with order).

Eq.\ (\ref{eq:P1order}) is not very practical for calculating the
thermal-power terms needed for evaluating the thermal impedance from the
temperature amplitudes obtained from Eq.\
(\ref{eq:LowestOrderSolution}), since $\Delta T$ is needed.

Alternatively the thermal power can be written as 
\begin{eqnarray}\label{eq:Palternative}
  P&=&I(U-V)=\frac{(U-V)V}{R_\text{pre}}
\end{eqnarray}
using Eq.\ (\ref{eq:IofVbasic}). From this equation the power can be
directly determined from measurable quantities. 

If only the $1\omegae$ component of the voltage is taken into account
the following approximations are obtained
\begin{multline}
  P\approx\frac{1}{4 R_\text{pre}}((U_1-V_1)E_1+c.c.)(V_1 E_1+c.c.)=\\
  \frac{1}{4 R_\text{pre}}((U_1-V_1) V_1^* +(U_1-V_1) V_1 E_2+c.c.)
\end{multline}
leading to power coefficient given as
\begin{subequations}\label{eq:Power1orderUV}
\begin{equation}\label{eq:Power0}
 P_0=\frac{1}{4 R_\text{pre}}((U_1-V_1) V_1^*+(U_1^*-V_1^*) V_1)
\end{equation}
and
\begin{equation}\label{eq:Power2}
 P_2=\frac{1}{2 R_\text{pre}}(U_1-V_1) V_1.
\end{equation}
\end{subequations}

A more general expansion including higher harmonics is given in appendix
\ref{app:power-terms-higher}.

\subsection{The AC-method refined}
\label{sec:ACrefined}
As mentioned earlier we have to do a refined analysis
of the detected harmonics of the voltages. These extra terms stem from
different sources: 1) The voltage source itself has higher harmonics
and a DC-offset. 2) A second order term in the temperature dependence
of the thermistor resistance. 3) The $4\omegae$ Joule-power component
produced by the $2\omegae$ variation of the thermistor (as discussed in
section \ref{sec:power-terms}). These effects are taken into account,
as described below.

Regarding point 1) we write the source voltage, $U(t)$, more generally
as
\begin{equation}\label{eq:Ufourthorder}
  U(t)=\frac{1}{2}\left(U_0 +U_1E_1+U_2E_2+U_3E_3+U_4E_4 + c.c. \right)
\end{equation}
including up to the fourth order harmonics.  A $U_3$ component from the
source is to be expected since the source itself has an inner
resistance. The $U_0$ is a DC-offset that was present in this
experiment but in principle should be eliminated.

Regarding point 2) we expand the temperature
dependence of the thermistor to second order as
\begin{equation}\label{eq:Rsecond}
R=R_0(1+\alpha_1\Delta T(t) + \alpha_2\Delta T(t)^2).
\end{equation}
Because $R(T)$ follows Eq.\ (\ref{eq:Rfull}) over an extended
temperature range $\alpha_2$ and $\alpha_1$ are connected as
\begin{equation}
  \alpha_2=\frac{\alpha_1^2}{2}-\frac{\alpha_1}{T_\text{cryo}}.
\end{equation}

Point 3) is taken into account by expanding  $V(t)$, $P(t)$ and
$T(t)$ like $U(t)$ in Eq.\ (\ref{eq:Ufourthorder}) up to the fourth
harmonic components also. 

The full analysis can be found in the appendix \ref{app:ACHigher},
including terms larger than $5\E{-7}U_1$.  Fortunately it leads to
rather simple extensions of the Eqs.  \ref{eq:LowestOrderSolution}
\begin{subequations}\label{eq:T0T2HighOrder}
\begin{eqnarray}
  A&=& \frac{\left[U_1-aT_0U_1-\frac{1}{2}aT_2U_1^*+X_1\right]}{V_1}-1  \label{eq:T0T2itera}\\
  T_2&=&-2\frac{V_3(A+1)-\left(U_3+X_3\right)}{aU_1}.
 \label{eq:T0T2iterb}
\end{eqnarray}
\end{subequations}
$X_1$ and $X_2$ depends on $T_0$, $T_1$, $T_2$, $T_4$, and
$b=\left(\frac{A\alpha_1}{1+A}\right)^2-\frac{A\alpha_2}{1+A}$, and
the expressions are given in Eq.\ (\ref{eq:V1V3_Full}).

The coupled Eqs.\ (\ref{eq:T0T2HighOrder}), can be solved by
iteration, using a procedure equivalent to the one for the first order
solution.  

We start by estimating $A(T)$ from Eq.\ (\ref{eq:T0T2itera}), putting
$X_1=T_0=T_2=0$, and from this estimate $\alpha_1$ and $\alpha_2$. 

A provisory thermal impedance $Z_2$ is then found from Eq.\
(\ref{eq:T0T2iterb}). $T_1=Z_1
P_1$ and $T_4=Z_4 P_4$ can then be estimated since
$Z_1(\omega)=Z_2(\omega/2)$ and $Z_4(\omega)=Z_2(2\omega)$, and $T_0$
is found from the limit of $Z_0$ as described earlier. 
The perturbative terms, $X_1,X_3$, can then be found and with these
values in Eqs.  (\ref{eq:T0T2itera}) and (\ref{eq:T0T2iterb}) a better
estimation of $A$ (and hence $\alpha_1$ and $\alpha_2$) and $T_2$ is
found.  This process is iterated until convergence.

The two important power components are calculated from the $V_i$'s
and $U_i$'s as in Eq.\ (\ref{eq:Power1orderUV}) by
\begin{equation}
 P_0 =\frac{1}{R_\text{pre}}\left( W_0V_0+\frac{1}{2}\Re\left(W_1V_1^* +
      W_2V_2^* + W_3V_3^* + W_4V_4^*\right)\right)
\end{equation}
and
\begin{multline}
 P_2  =\frac{1}{R_\text{pre}}\left(
    W_0V_2+W_2V_0+ \vphantom{\frac{1}{2}}  \right. \\ \left.
    \frac{1}{2}\left(W_1 V_1+W_1^*V_3+W_2^*V_4+W_3 V_1^*+W_4 V_2^*
    \right)\right)
\end{multline}
where $W=U-V$. The full set of power components are given in Eq.\
(\ref{eq:FullPs}).

\section{Experimental}
\label{sec:expermental}
The measurements were performed using a custom built setup
\cite{Igarashi2008a,Igarashi2008b}. The temperature was controlled by a
cryostat with temperature fluctuations smaller than
$2\milli\kelvin$ (see Ref. \onlinecite{Igarashi2008a} for details on
the cryostat).  The electrical signals were measured using a HP3458A
multimeter in connection with a custom-built frequency generator as
sketched on Fig.\ \ref{fig:vdiv} (see Ref.\
\cite{Igarashi2008b} for details on the electrical setup, but
notice that in this experiment we used the generator and multimeter in
a slightly altered configuration). The electrical setup allows for
measurements in the frequency ($\nue$) range from $1\milli\hertz$ up to
$100\hertz$.

\begin{figure}
  \centering
  \includegraphics[width=8.6cm]{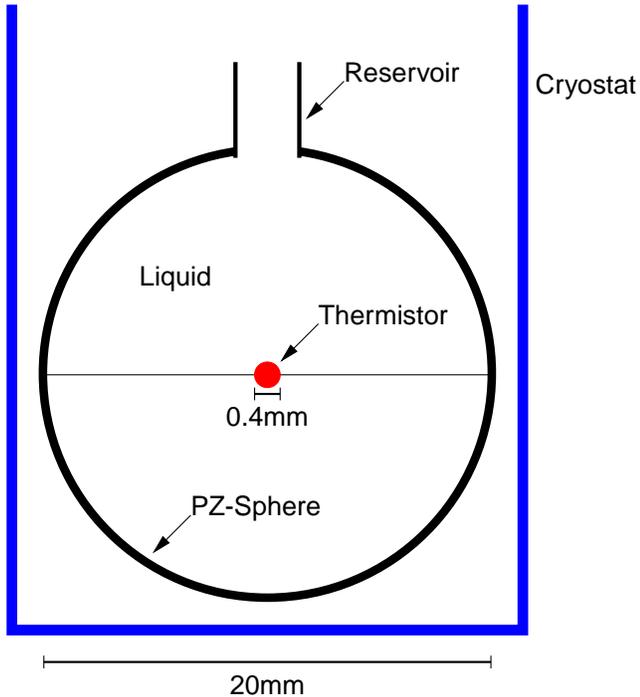}  
  \caption{Schematic illustration of the measuring setup. The
    thermistor acting as combined heater and thermometer is placed in
    a sphere filled with the liquid. The sphere is itself placed in a
    cryostat.}
  \label{fig:BeadInSphere}
\end{figure}

The used thermistor was of the type U23UD from Bowthorpe
Thermometrics, and was positioned in the middle of a sphere as shown
on Fig.\ \ref{fig:BeadInSphere}.  The details of the sphere is as
discussed earlier unimportant, and in this case the piezoelectric bulk
transducer \cite{Chr94} is used as such.  The sphere is
filled at room temperature with additional liquid in the reservoir.
This additional liquid ensures that the sphere stays completely filled as
it is cooled to the relevant temperature.

The liquid investigated was a five-ring polyphenyl ether (5-polyphenyl-4-ether)
the Santovac\textregistered{} 5 vacuum pump fluid (CAS number 2455-71-2).
The liquid was used as received, without further purification.

Data were taken as frequency scans at constant temperature, with the
liquid in thermal equilibrium.  The direct output from the frequency
generator was measured under electrical load, at each sample
temperature.  This eliminates any long term drifts in the frequency
generator output voltage. The measurements were furthermore carried
out at two different input amplitudes, $\langle|U_1|\rangle\approx
4.9\volt$ and $\langle|U_1|\rangle\approx 2.9\volt$, allowing for a
test of the inversion algorithm and linearity of the final thermal
response. If not stated otherwise all results presented in this paper
are from the larger of the two amplitudes.  To test for
reproducibility and equilibrium some of the measurements were retaken
during the reheating of the sample.  The outcome of the measurements
are the complex amplitudes $U_i(\nue)$ and $V_i(\nue)$ for the
harmonics of the voltages $U$ and $V$.

Table \ref{tab:ExpParm} gives the key experimental parameters for the
measurements at the high input amplitude.

\section{Data analysis}
\label{sec:Deep-analysis}
In the following we denote the measured thermal impedance, $Z_2$, by
simply $Z$ dropping the subscript $2$. The subscripts were practical
in the technical description of the detection method indicating that
the thermal impedance is found at a thermal frequency, $\omega$, which
is the double of the electrical frequency, $\omega_e$. But in
discussing the thermal properties of liquids only the thermal
frequency $\omega=2 \pi \nu$ is relevant. The temperature we refer to
at which a measurement is done is the mean temperature at the bead
surface, $T=T_0+T_\text{cryo}$.

The thermal impedance has been measured at two different
input voltage amplitudes, $U_1$, of the voltage divider of Fig.\
\ref{fig:vdiv} and thus at two different power amplitudes. The
liquid thermal response, the temperature, is expected to be linear in
the power amplitudes and thus the thermal impedance should be
independent of the power amplitude. 

The thermal impedance has been calculated from the measured voltage
harmonics using both the first-order method described in section
\ref{sec:first-order-model} and using the higher-order method
described in section \ref{sec:ACrefined}. 

\begin{figure}
  \centering
  \includegraphics[width=8.6cm]{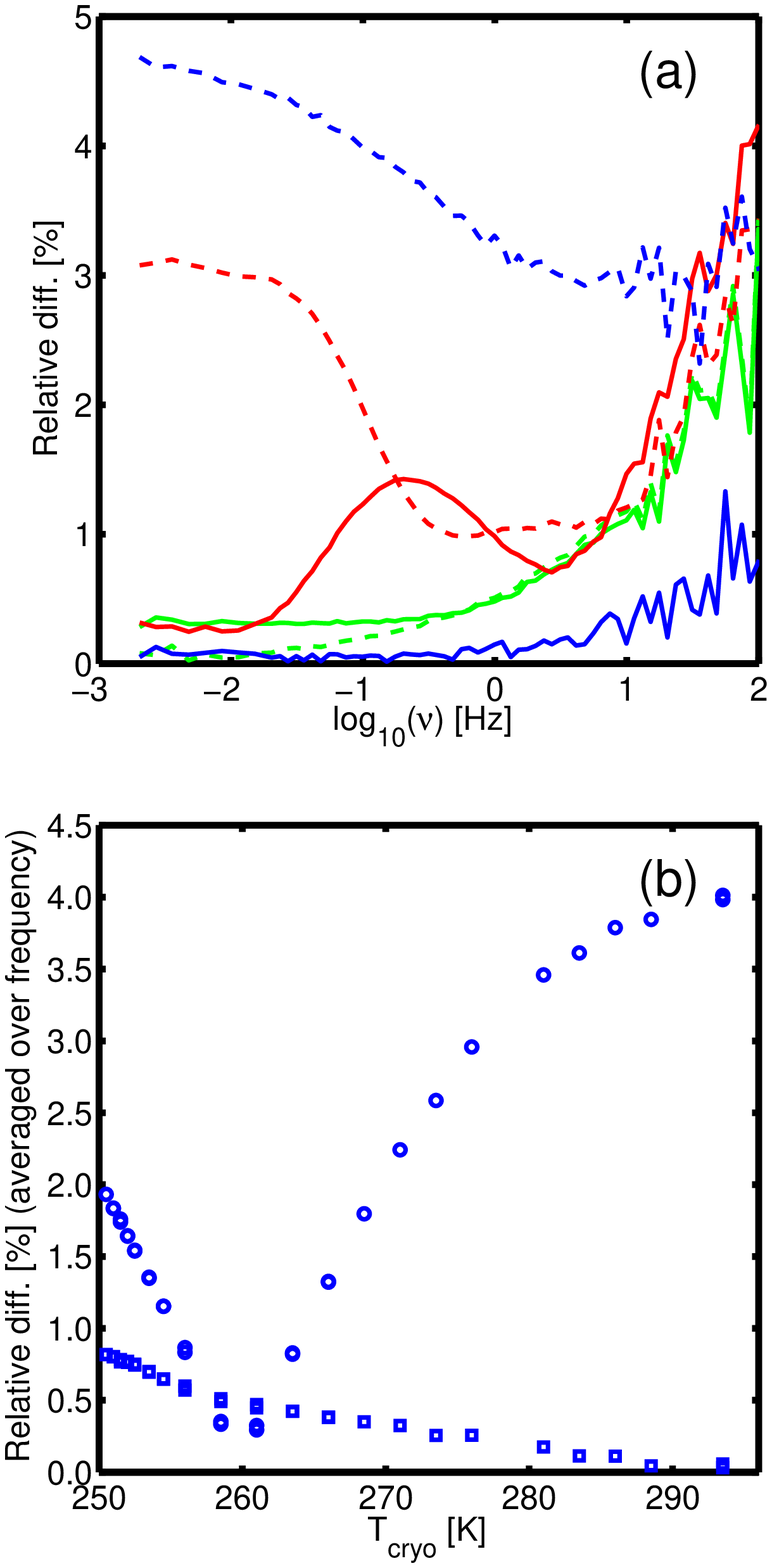}  
  \caption{Comparison of the thermal impedance, $Z$, obtained from the
    measurements at two different amplitudes ($\langle U_1 \rangle =
    4.9\volt$ and $2.9\volt$), and from using the first-order model as
    given in Eq.\ (\ref{eq:LowestOrderSolution}) and higher-order
    model as given in Eq.\ (\ref{eq:T0T2HighOrder}) (as described in
    section \ref{sec:first-order-model} and \ref{sec:ACrefined}).\\
    \textbf{(A)} the frequency dependent relative
    difference between the two methods
    ($|Z_{\text{amp}=2.9\volt}-Z_{\text{amp}=4.9\volt}|/|Z_{\text{amp}=2.9\volt}|$).
    \textbf{Full lines:} the difference for the higher-order model.
    \textbf{Dashed lines:} the difference for the first-order model.
    Cryostat temperatures are $293.5\kelvin$ (blue), $261\kelvin$
    (green) and  $250.5\kelvin$ (red). \\
    \textbf{(B)} shows the frequency averaged relative
    difference
    ($\langle|Z_{\text{amp}=2.9\volt}-Z_{\text{amp}=4.9\volt}|\rangle_{\nu}/\langle|Z_{\text{amp}=2.9\volt}|\rangle_{\nu}$). \textbf{Squares}
    the difference for the higher-order model.  \textbf{Circles} the
    difference for the first-order model. }
  \label{fig:ZthAt2Amps}
\end{figure}

The top part of Fig.\ \ref{fig:ZthAt2Amps} shows the frequency
dependent relative difference between the calculated thermal impedance
from the two amplitudes. The results from the two methods (the
first-order and higher-order methods) for three selected temperatures
are shown in the figure. In the lower part the average relative
difference is shown as function of temperature. It is seen that the
higher order method generally reduces the difference significantly.
Only in a very small temperature interval does the two methods give
equally good results. This demonstrates the necessity of
taking into account the higher-order terms of the elaborated analysis
(as described in section \ref{sec:ACrefined}).

\begin{figure}
  \centering
  \includegraphics[width=8.6cm]{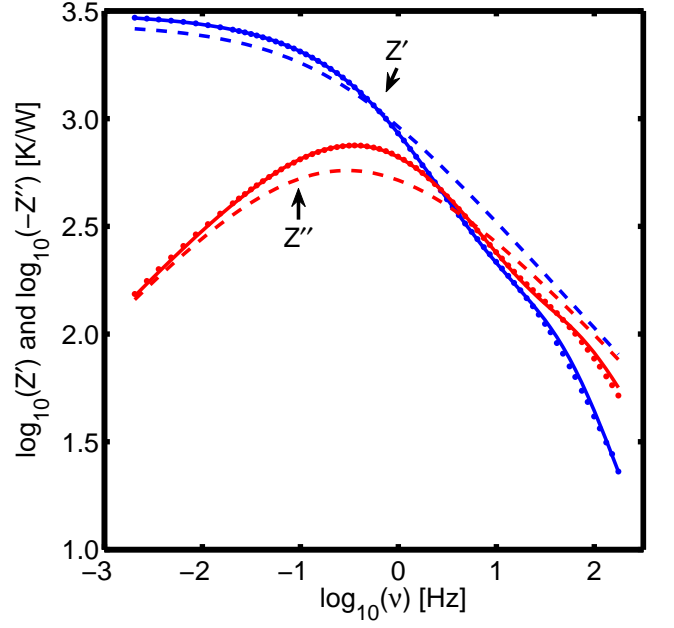}  
  \caption{Comparison of measured thermal impedance, $Z$, with the
    result from fitting the full model of liquid and thermistor (Eqs.\
    (\ref{eq:TESG2}), (\ref{eq:Zr0(Zliq)}) and (\ref{eq:Z(Zr0)})), and
    the simple model of $Z_\text{liq}$ (Eq.\ (\ref{eq:Zliq5parm})).
    The parameters used in evaluation of the simple model are the ones
    found from the full analysis. Data are taken at room temperature
    ($T=295.6\kelvin$).  \textbf{Dots:} measured points. \textbf{Full
      lines:} full model of liquid and thermistor bead. \textbf{Dashed
      lines} simple model including only the liquid.}
  \label{fig:SimpelAndFullModelCompared}
\end{figure}

\begin{figure}
  \centering
  \includegraphics[width=8.6cm]{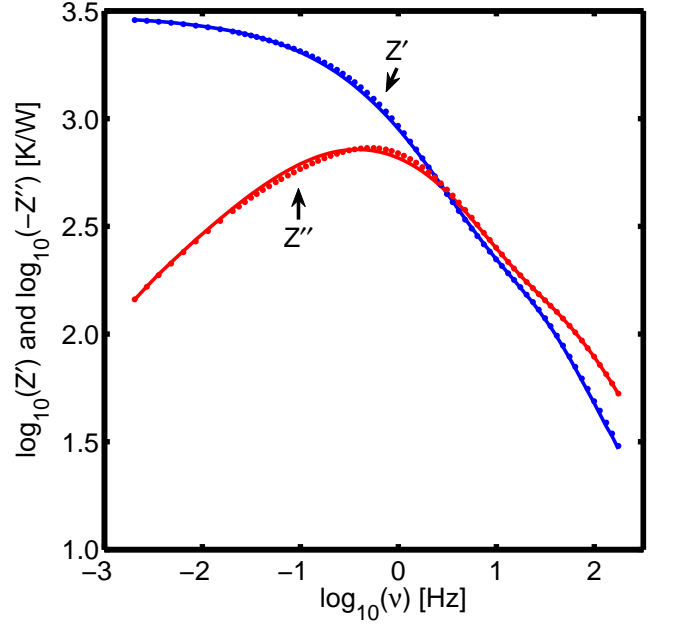}  
  \caption{Comparison of thermal impedance, $Z$, at two temperatures,
    showing the small difference due to the frequency dependent
    specific heat of the liquid, $c_l(\nu)$.  \textbf{Lines:}
    $T=256.9\kelvin$. \textbf{Dots} $T=252.8\kelvin$.}
  \label{fig:ZthAtTwoTemps}
\end{figure}

Fig.\ \ref{fig:SimpelAndFullModelCompared} shows the real and
imaginary part of the thermal impedance, $Z$, in a log-log plot at
$295.6\kelvin$. The measured impedance has more features than
predicted by the simple expression for the thermal impedance of
spherical effusion Eq.\ (\ref{eq:TESG2}). For example the high
frequency behavior should be characterized by a line of slope $1/2$
but this is seen not to be the case. The additional features come from
thermal properties of the thermistor bead, the influence of which we
are going to study below. We stress that although $Z$ shows dispersion
and has a real and imaginary part this has nothing to do with liquid
relaxation --- which is absent at this temperature --- but is only a
consequence of heat diffusion in spherical geometry. However
relaxation should of course affect the thermal impedance. This is
shown in Fig.\ \ref{fig:ZthAtTwoTemps} where $Z$ at $252.7\kelvin$ and
$256.7\kelvin$ are compared. The glass transition is seen to give only
a slight perturbation of the shape of the $Z$-curve. In order to
obtain any reliable specific heat data from the thermal impedance it
is thus imperative to have an accurate model of the thermal
interaction between liquid and thermistor bead.  

\begin{figure}
   \centering
   \includegraphics[width=8.6cm]{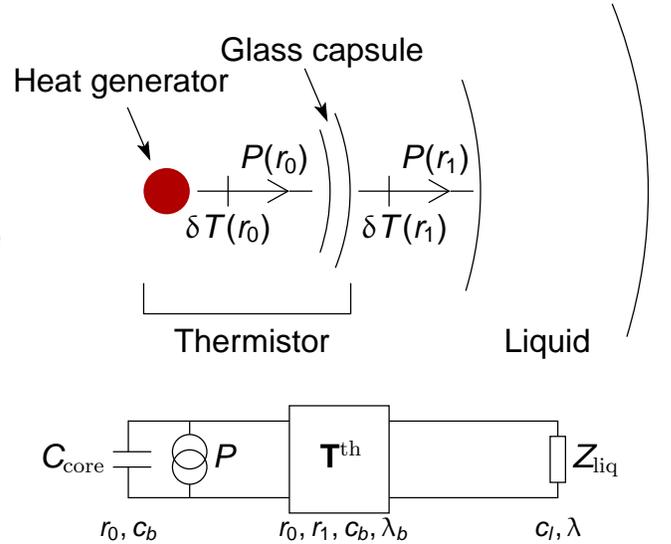} 
   \caption{Schematic thermal network model of the thermistor bead and
     liquid. The thermal current $P$ and temperature $\delta T$ are
     indicated at the inner radius, $r_0$, at the border between the
     heater and the glass capsule and at the outer radius, $r_1$, at
     the border between the glass capsule and the liquid. The thermal
     impedance of the liquid is defined as $Z_\text{liq}=\delta
     T(r_1)/P(r_1)$. The thermal impedance seen from the thermistor is
     $Z=\delta T(r_0)/P$. The two are connected by Eq.\ (\ref{eq:Zr0(Zliq)})
      and (\ref{eq:Z(Zr0)}), utilizing the thermal
     transfermatrix Eq.\ (\ref{eq:TrmthFull5parm}). }
  \label{fig:bead}
\end{figure}

The bead (see Fig.\ \ref{fig:bead}) consists of a core of radius $r_0$,
where the temperature amplitude $\delta T$ is actually measured
and the heat current $P$ generated. This core is surrounded by a
glass capsule of outer radius $r_1$, at which the bead is in contact
with the liquid.  The heat flow $P(r_1)$ out through the surface at
$r_1$ is in general different from the heat flow $P(r_0)$ out through 
the surface at $r_0$. Also the temperature $\delta T(r_1)$ can be different 
from $\delta T(r_0)$. Now the liquid thermal impedance is
\begin{equation}\label{eq:Zr1}
 Z_\text{liq}=\frac{\delta T(r_1)}{P(r_1)}
\end{equation}
whereas the measured thermal impedance is rather
\begin{equation}\label{eq:Zr0}
 Z=\frac{\delta T}{P}.
\end{equation}
The capsule layer has a heat conductivity $\lambda_b$ and a specific
heat $c_b$. The bead is a solid for which in general the difference between the isobaric and
isochoric specific heats are small and they are frequency independent. The
heat diffusion through this shell is thus well described \cite{chr08b}
by a thermal transfermatrix, $\textbf{T}^{\rm th}=\textbf{T}^{\rm
  th}(\lambda_b,c_b,r_1,r_0)$,
\begin{equation}\label{eq:Trmth}
\begin{pmatrix}
\delta T(r_1)\\
P(r_1)/(i\omega)
\end{pmatrix}\,=\,
\textbf{T}^{\rm th}
\begin{pmatrix}
  \delta T(r_0)\\
  P(r_0)/(i\omega)
\end{pmatrix}\,.
\end{equation}

This implies that the thermal impedance $Z_\text{liq}=Z_{r_1}$ at radius $r_1$ is transformed to a thermal impedance $Z_{r_0}$ at radius $r_0$ via
\begin{equation}\label{eq:Zr0(Zliq)}
  Z_{r_0}=\frac{1}{i\omega}\frac{T^{\rm th}_{12} - T^{\rm th}_{22} i\omega
    Z_\text{liq}}{ T^{\rm th}_{11}-T^{\rm th}_{21}i\omega
    Z_\text{liq}}.
\end{equation}
A small part of generated power is stored in the core of the bead. The heat capacity $C_\text{core}$ of the core is taken to be $4/3 \pi {r_0}^3 c_b$. The measured thermal impedance $Z$ is thus related to $Z_{r_0}$ by
\begin{equation}\label{eq:Z(Zr0)}
 \frac{1}{Z}=i\omega C_\text{core}+\frac{1}{Z_{r_0}}
\end{equation}
The model relating $Z_\text{liq}$ to the measured $Z$ described by Eqs. (\ref{eq:TESG2}), (\ref{eq:Zr0(Zliq)}) and (\ref{eq:Z(Zr0)}) are depicted in the electric equivalent diagram in Fig \ref{fig:bead}.

In order to get $Z_\text{liq}$ from the measured $Z$ one has to know
the parameters $r_0,r_1,\lambda_b$ and $c_b$ characterizing the
thermal structure of the intervening thermistor bead. 

This can in principle be done by fitting $Z$ of the model
obtained by combining Eqs.\ \eqref{eq:TESG2}, \eqref{eq:Zr0(Zliq)}, and
\eqref{eq:Z(Zr0)} to the measured frequency dependent thermal
impedance. An electric equivalent network of the model is depicted in Fig.\ \ref{fig:bead}.
This model has six frequency independent intrinsic
parameters, namely $r_0,r_1,\lambda_b,c_b,\lambda$ and $c_l$, if the
fit is limited to the temperature region where $c_l$ can be considered
frequency independent. By careful inspection of the equations it can
be found that the number of independent parameters that can be deduced from a fit is only five. The
five parameters can be chosen in a number of ways, of which we have chosen 
\begin{multline}
  \label{eq:4}
  \tau_l=r_1^2\frac{c_l}{\lambda},\ 
  \tau_b=r_1^2\frac{c_b}{\lambda_b},\  \hat{r}=\frac{r_0}{r_1},
  \quad \hat{c}=\frac{c_l}{c_b} ,\ Z_{\text{liq},0}=\frac{1}{4\pi\lambda r_1}.
\end{multline}
This set of variables has the advantage that it minimizes the number
of parameters which mix bead and liquid properties. Here $\hat{c}$
is the only one such mixed parameter. The expressions for the transfermatrix and $Z_\text{liq}$ in terms of these parameters is given in appendix \ref{sec:full-five-parameter}.

\begin{figure}
  \centering
  \includegraphics[width=8.6cm]{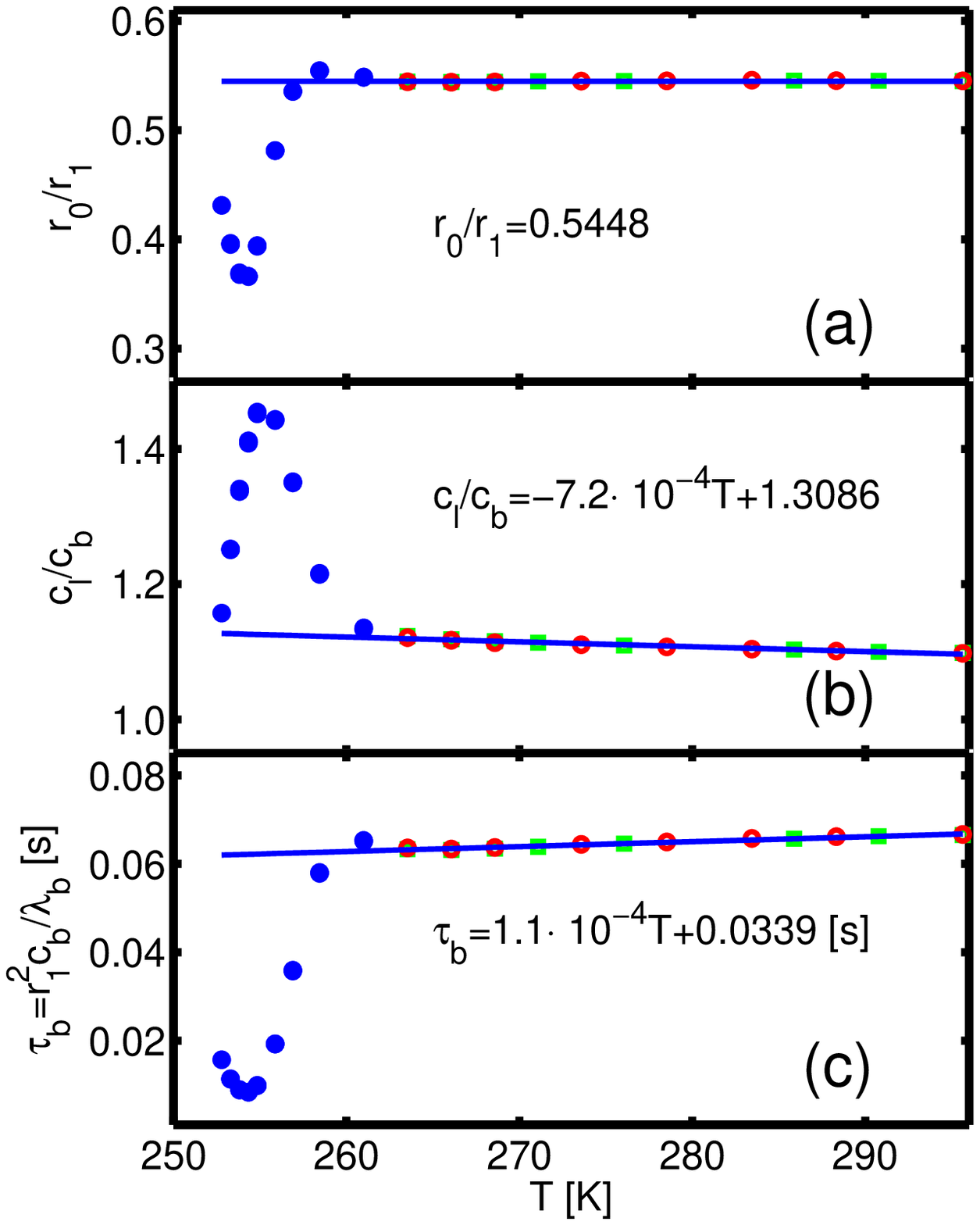}  
  \caption{Results from fitting the measured thermal impedance, $Z$,
    to the $5$-parameter thermal network model (Fig.\ \ref{fig:bead})
    of the spherical heat effusion experiment Eqs.\ (\ref{eq:TESG2}), (\ref{eq:Zr0(Zliq)}) and (\ref{eq:Z(Zr0)}).
    This figure shows (a) $\hat r=r_0/r_1$ , (b) $\hat
    c=c_l/c_b$ and (c) $\tau_b$, (the remaining parameters
    are shown on figure \ref{fig:tauZ}). \\ \textbf{Open squares
      (red):} Data taken going down in temperature. \textbf{Open
      circles (green)}: Data taken going up in
    temperature. \textbf{Closed circles (blue):} Low temperature data
    including points take going down and up.  \textbf{Line:}
    Interpolated temperature dependencies of $\hat r$, $\hat
    c$ and $\tau_b$ found based on data from above the
    liquid glass transition temperature range, used for extrapolating
    down in the region where the liquid relaxation sets in (the
    functional form and fitting results are shown).}
  \label{fig:rctaub}
\end{figure}

\begin{figure}
  \centering
  \includegraphics[width=8.6cm]{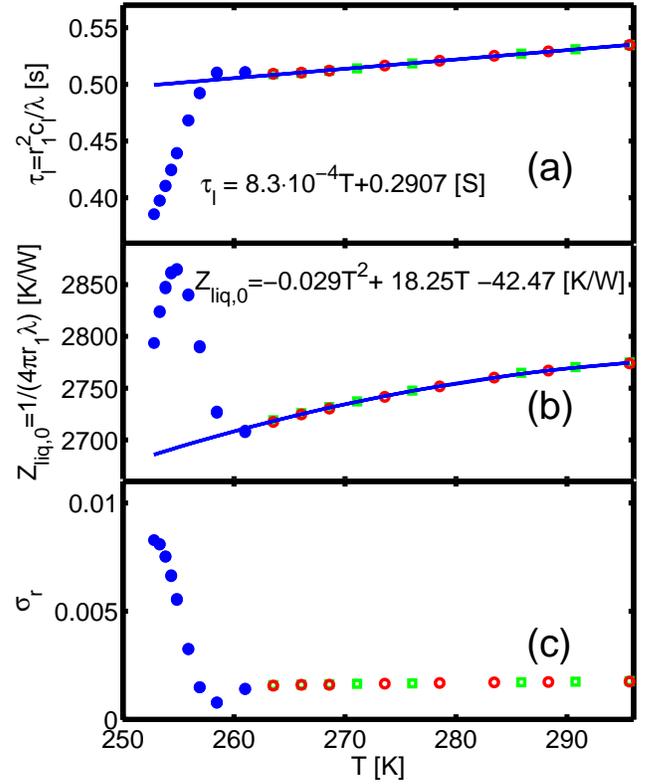}  
  \caption{Results from fitting the measured thermal impedance, $Z$,
    to the $5$-parameter thermal network model (Fig.\ \ref{fig:bead})
    of the spherical heat effusion experiment Eqs.\ (\ref{eq:TESG2}), (\ref{eq:Zr0(Zliq)}) and (\ref{eq:Z(Zr0)}).
    This figure shows (a) $\tau_l$, (b) $Z_{\text{liq},0}$, and (c) the relative
    error from the fit,
    $\sigma_r=\sqrt{\frac{\sum_{\nu}|Z-Z_\text{model}|^2}{\sum_{\nu}|Z_\text{model}|^2}}$),
    (the  remaining parameteres are shown on figure \ref{fig:rctaub})\\
    \textbf{Open squares (red):} Data taken going down in
    temperature. \textbf{Open circles (green)}: Data taken going up in
    temperature. \textbf{Closed circles (blue):} Low temperature data
    including points take going down and up.  \textbf{Line:}
    Interpolated temperature dependencies of $\tau_l$, and $Z_{\text{liq},0}$ found
    based on data from above the liquid glass transition temperature
    range, used for extrapolating down in the region where the liquid
    relaxation sets in (the functional form and fitting results are
    shown).}
  \label{fig:tauZ}
\end{figure}

The five parameter model was fitted to the measured thermal impedance,
$Z$, by using a least square algorithm. The results are shown on Fig.\
\ref{fig:rctaub} and \ref{fig:tauZ}. Measurements were taken both
going downwards to the glass transition and upwards again to room
temperature. In the glass transition range where the liquid specific
heat, $c_l$, is expected to be complex and frequency dependent these
fits of course do not work. This is the reason for the scatter of
data at low temperature. This difference between the high and low
temperature regimes can also be seen directly in the error from the
fit as illustrated on the lower part of Fig.\ \ref{fig:tauZ}.

The solid line in Fig.\ \ref{fig:SimpelAndFullModelCompared} shows the
result of such a fit to $Z$-data at room temperature ($295.6$K). The
model captures the features in the frequency dependence of the
measured thermal impedance very precisely. 
Also in Fig.\ \ref{fig:SimpelAndFullModelCompared} is shown the
thermal impedance as it would have been if the properties of the
thermistor bead itself had been negligible, i.e. if $Z$ was identical
to $Z_\text{liq}$ of Eq.\ (\ref{eq:TESG2}) We note that especially the
high frequency behaviors are different but also the limiting value at
low frequency is higher for the measured $Z$ than for
$Z_\text{liq}$. This is due to the additional thermal resistance in
the bead capsule layer.

In table \ref{tab:FitPamnRepro} we report the results for $\tau_b$ and
$\hat c$ obtained at room temperature. Four values are given for
each quantity, they are taken at the two input amplitudes and from an
initial measurement at room temperature and a measurement taken at the
end of the temperature scan. The method is seen to give very
reproducible results.

\begin{table*}
  \centering
  \caption{Heat conductivity and specific heat
    for the liquid ($\lambda$ and $c_l$), as defined in relation to Fig.\
    \ref{fig:Lliq} and \ref{fig:Czero}. Fitting parameters
    characterizing the bead ($\tau_b$ and $\hat c$) from the fits to
    the five-parameter mode.
    The results are at room temperature ($295.6$K), and taken using
    two input amplitudes $|U_1|$. The two scans are the initial scan
    and a scan taken after the temperature scan coming back to room
    temperature. The results  indicates the reproducibility of the method. 
  }
  \label{tab:FitPamnRepro}
  \setlength{\tabcolsep}{0.7em}
   \begin{tabular}{ccccccc}
   \hline \hline
 $|U_1|$   & $P_2$ & scan  & $\tau_b$ & $\hat c=c_{l}/c_b$  & $\lambda$  & $c_l$\\ 
 $[\volt]$ & $[\milli\watt]$ & & $[s]$ & & $[\watt/(\kelvin\cdot\meter)]$ & $[10^6
 \joule/(\kelvin\cdot\meter^3)]$   \\ \hline
 \raisebox{-1.5ex}[0pt]{$4.9$}  &\raisebox{-1.5ex}[0pt]{$0.63$} & $1$ &  $ 0.0665$  & $1.0983$ & $0.13759$   & $1.7597$\\
                                &                               & $2$ &  $ 0.0666$  & $1.0977$ & $0.13761$   & $1.7602$\\ \hline
 \raisebox{-1.5ex}[0pt]{$2.9$}  &\raisebox{-1.5ex}[0pt]{$0.23$} & $1$ &  $ 0.0663$  & $1.0989$ & $0.13786$   & $1.7627$\\
                                &                               & $2$ &  $ 0.0668$  & $1.0953$ & $0.13782$   & $1.7623$\\ 
   \hline \hline
  \end{tabular}
\end{table*}

The five parameters are extrapolated down to the regime of frequency
dependent $c_l$, by the functions indicated on the figures. $\hat r$ is
considered being temperature independent. $\hat c$, $\tau$ and $\tau_b$ are to
a good approximation linear in temperature over the investigated
temperature range. $Z_{\text{liq},0}$ is found to be very well fitted by a second
degree polynomial.

\begin{figure}
  \centering
    \includegraphics[width=8.6cm]{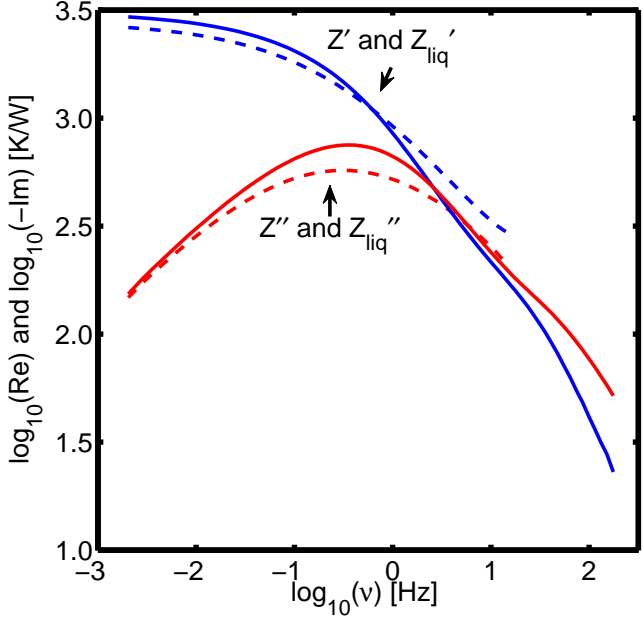}  
    \caption{Comparison of the raw measured thermal impedance, $Z$,
      and final liquid thermal impedance, $Z_\text{liq}$, corrected
      for the influence of the thermistor encapsulation.  Data are
      taken at room temperature ($T=295.6\kelvin$), and are equivalent
      to the data shown on Fig.\ \ref{fig:SimpelAndFullModelCompared}.
      The liquid thermal impedance is truncated at high frequencies
      ($\nu>10^{1.2}\hertz$) as the bead becomes dominant at higher
      frequencies, leading to unphysical behavior of the calculated
      liquid impedance. \textbf{Full lines:} Raw thermal impedance
      $Z$. \textbf{Dashed lines:} Liquid thermal impedance $Z_\text{liq}$.
      }
  \label{fig:Zliq}
\end{figure}

The extrapolated values of the five parameters can then be used for calculating
$Z_\text{liq}$ from the measured thermal impedance, $Z$ inverting the relations 
Eqs.\ (\ref{eq:Z(Zr0)}) and (\ref{eq:Zr0(Zliq)}). In Fig.\
\ref{fig:Zliq} the thermal impedance of the liquid, $Z_\text{liq}$,
and the raw measured thermal impedance, $Z$, are shown for data taken
at room temperature ($=295.6\kelvin$).

As discussed earlier effusion from a plane plate can only give the
effusivity $e=\sqrt{\lambda c}$. Thus the specific heat cannot be
found absolutely. In contrast effusion from a sphere is able to give
both the heat conductivity and the longitudinal specific heat
absolutely even if the latter is frequency dependent. However, the heat
conductivity has to be frequency independent to allow for this
separation. The separation is possible because $Z_\text{liq}$ as given
in Eq.\ (\ref{eq:TESG2}) has the low frequency limiting value
$Z_{\text{liq},0}=1/(4\pi\lambda r_1)$. The convergence is rather slow involving
the squareroot of the frequency. However, if we look at the reciprocal
quantity, the thermal admittance $Y_\text{liq}=1/Z_\text{liq}$
\begin{equation}
Y_{\text{liq}}=4 \pi \lambda r_1 \left(1+ \sqrt{i\omega {r_1}^2 c_l/\lambda}\right)
\end{equation}
we observe that when $c_l$ is frequency independent the squareroot
frequency term can be cancelled by taking the difference between
$Y_\text{liq}'$ and $Y_\text{liq}''$
\begin{equation}\label{eq:ReYliqminusImYliq}
\lambda =\frac{Y_\text{liq}'-Y_\text{liq}''}{4 \pi r_1}.
\end{equation}
This is valid even in the glass transition range at sufficiently low
frequencies where $c_l$ reaches its non-complex equilibrium value.
Finally, using the found values for $\lambda$ the frequency
dependent longitudinal specific heat is calculated from
\begin{equation}
 c_l(\omega)=\frac{\lambda}{i\omega r_1^2}\left(\frac{Y_\text{liq}}{4 \pi \lambda r_1}-1\right)^2
\end{equation}

However the structure of the five parameter model only allows for determining
$\hat r$ but not $r_0$ and $r_1$ independently from the fit to data. To
obtain the absolute value of $\lambda$ and $c_l$ one of the radii are
needed.  Measured values of $r_1$ are given in table
\ref{tab:ExpParm}. As indicated the bead is not completely spherical,
and it is therefore needed to determine which value of $r_1$ to use.  

According to the specifications for Santovac\textregistered{} 5 given
by Scientific instrument service, Inc \cite{sis} the value of
$\lambda$ at $20\celsius$ is $0.1330\watt/(\kelvin\cdot\meter)$. They
likewise report the isobaric specific heat pr.\ mass
$c_p/\rho=0.35\text{cal}/(\gram\cdot\kelvin)$ and density
$\rho=1.204\gram/\centi\meter^3$ at $20\celsius$ giving a specific
heat pr. volume of $c_p=1.76\E{6}\joule/(\kelvin\cdot\meter^3)$.  At
these temperatures the isobaric and longitudinal specific heats are
identical, we have therefore chosen to use a value for $r_1$ which
matches $c_l$, within the errors, to the reference value. This
``optimized'' value is very close to the $r_{1,\text{Measured, long
    edge}}$ value given in table \ref{tab:ExpParm}. This choice gives
a value of $\lambda$ which is $3.5\%$ too big compared to the literature
value.

\begin{figure}
  \centering
  \includegraphics[width=8.6cm]{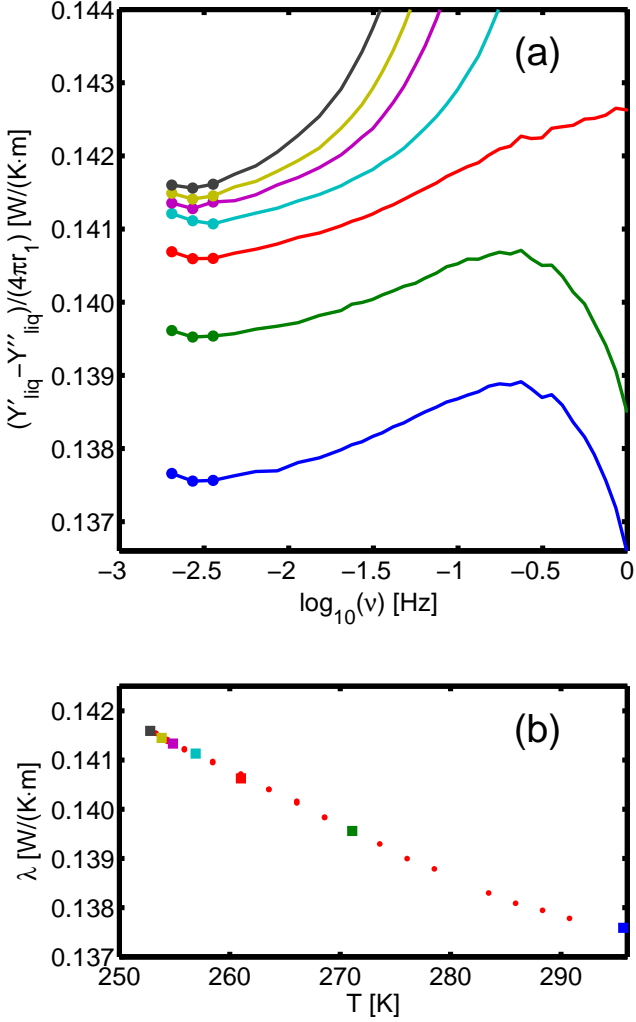}  
  \caption{Thermal conductivity of the liquid. \textbf{(a):}
    $\frac{Y'_\text{liq}-Y''_\text{liq}}{4\pi r_1}$ as function of
    frequency, plotted for a selection of temperatures (same
    temperatures as on Fig.\ \ref{fig:Cl}, with colors indicating same
    datasets). According to Eq.\ \ref{eq:ReYliqminusImYliq} the
    thermal conductivity is the low frequency limit of this quantity.
    The points on the lines indicate the data points used for defining
    the low frequency limit (taken as an average over these points).
    \textbf{(b):} Thermal conductivity of the liquid defined from
    the average over the low frequency limit of
    $\frac{Y'_\text{liq}-Y''_\text{liq}}{4\pi r_1}$.  Squares
    corresponds to the curves shown on the upper figure, and dots to
    the remaining datasets.}
  \label{fig:Lliq}
\end{figure}

\begin{figure}
  \centering
    \includegraphics[width=8.6cm]{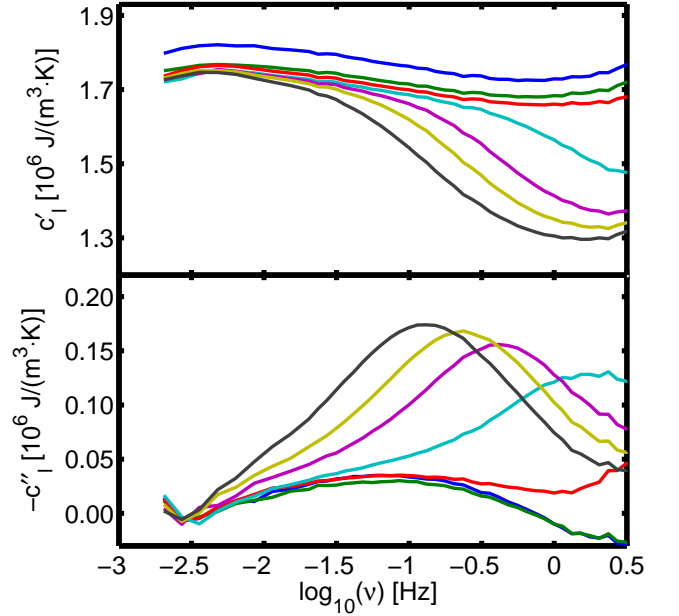} \\
    \caption{Real and imaginary part of the longitudinal specific heat
      of the liquid, $c_l$, at a number of the investigated
      temperatures. The frequency axis is truncated at $0.5\hertz$ as
      the signal above this frequency is dominated by the
      inner-structure of the thermistor. Temperatures are
$295.6\kelvin$,  $271.1\kelvin$, $261.0\kelvin$, $256.9\kelvin$, $254.9\kelvin$, $253.8\kelvin$, and $252.8\kelvin$.}
  \label{fig:Cl}
\end{figure}

Fig.\ \ref{fig:Lliq} shows how the temperature dependence of the heat
conductivity, $\lambda$, of the sample liquid is found. In Fig.\
\ref{fig:Cl} is shown the final frequency dependent specific heat for a number
of temperatures. Table \ref{tab:FitPamnRepro} reports the room
temperature values for $c_l$ and $\lambda$ for two temperature
amplitudes, and data reproduced after a temperature scan down and up again.
The values indicate the reproducibility of the method. 

\section{Discussion}
\begin{figure}
  \centering
   \includegraphics[width=8.6cm]{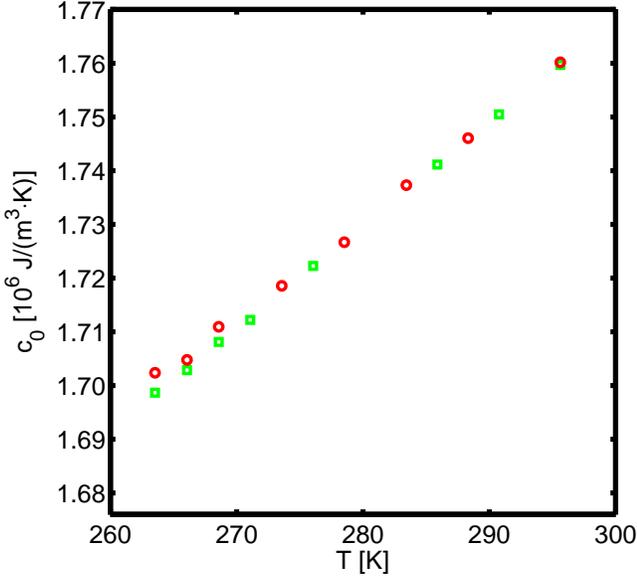} \\
    \caption{DC specific heat, $c_0$, of the liquid for temperatures
      above the liquid glass transition temperature range. The
      specific heat is found as the average over the real part of $c_l$ in
      the frequency range $10^{-2}\hertz$--$1\hertz$. $c_0$ 
      corresponds to the DC value of $c_p$ as can be seen from Eq.\
      (\ref{eq:cldeff}) and Eq.\ (\ref{eq:cpdeff}).  \textbf{Open
        squares (green):} Data taken going down in temperature. \textbf{Open
        circles (red)}: Data taken going up in temperature.
    }
  \label{fig:Czero}
\end{figure}

\begin{figure}
  \centering
   \includegraphics[width=8.6cm]{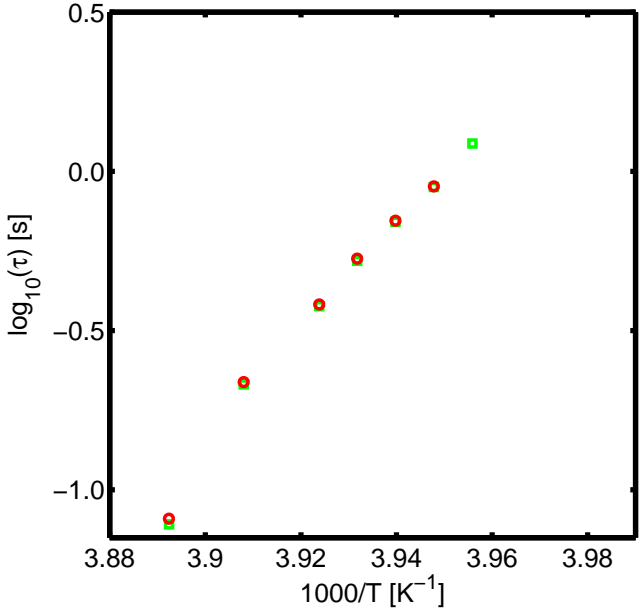} \\
    \caption{Characteristic relaxation time, $\tau$, for the
      longitudinal specific heat, $c_l(\omega)$. The time if found
      from a fit to 2.order polynomial to a few point around the
      maximum of the imaginary part of $c_l$ and  corresponds to
      $\tau=1/(2\pi\nu_\text{lp})$, where $\nu_\text{lp}$ is the
      frequency of maximum loss. \textbf{Open squares (green):} Data taken
      going down in temperature. \textbf{Open circles (red)}: Data taken
      going up in temperature. 
    }
  \label{fig:tau}
\end{figure}

The derived longitudinal specific heat displayed in Fig.\ \ref{fig:Cl}
shows the expected relaxation phenomena at the glass transition. At
$252.7\kelvin$ $c_l$ decreases from $c_0=c_l(\omega\rightarrow 0)=1.75
\cdot 10^6 \joule / (\kelvin \cdot {\meter}^3)$ to
$c_\infty=c_l(\omega\rightarrow \infty)=1.30 \cdot 10^6 \joule /
(\kelvin \cdot {\meter}^3)$. The temperature dependency of $c_0$ above
the glass transition range is shown in Fig.\ \ref{fig:Czero}.  The
temperature dependency of the characteristic relaxation time, $\tau$,
is shown in Fig.\ \ref{fig:tau}.  The high temperature data are seen
to show spurious frequency dependency in both the real and imaginary
part of $c_l$. This may be caused by small systematic errors in
the data processing or data acquisition that we have not been able to
trace yet. It may also be due to the small eccentricity of the bead.
Due to the functional relationship, Eq.\ (\ref{eq:TESG2})
between $Z_{\text{liq}}$ and $c_l$ a small relative error
$|dZ_\text{liq}/Z_\text{liq}|$ in the thermal impedance may propagate
to a large relative error $|dc_l/c_l|$ in the specific heat. As a
function of the dimensionless Laplace-frequency $s=i\omega \tau_l$ one
finds
\begin{equation}
 \frac{dc_l}{c_l}/\frac{dZ_\text{liq}}{Z_\text{liq}}=2\frac{1+\sqrt{s}}{\sqrt{s}}.
\end{equation}
This factor is four at the characteristic heat diffusion frequency
($s=1$) increasing to a factor of $22$ at $0.01$ times this frequency.
Thus the specific heat is difficult to get reliable more than 2
decades below the characteristic diffusion frequency. On the other
hand one decade above this frequency the measured $Z$ is dominated by
the thermal structure of the thermistor bead that can only to some
extent be reliable modelled .

\section{Conclusion}
From a theoretical point of view determination of the frequency
dependent specific heat by a thermal effusion method based on a sphere
has many ideal features. The surface of a sphere --- being closed ---
has no boundary and thus no associated boundary effects like a finite
plane plate. The thermomechanical coupling problem can be treated
analytically i.e. the influence of the increasing dynamic shear
modulus and the mechanical boundary conditions are adequately taken
into account. Although the thermomechanical problem can also be solved
in the onedimensional unilateral case it is not so obvious whether it
really applies to the experimental finite plate realizations. It is of
course preferable if thermal experiments can give two independent
thermal properties e.g. the specific heat and the heat conductivity or
the effusivity and diffusivity. Effusion experiments in different
geometries give the effusivity only unless a characteristic length
scale comes into play with the heat diffusion length. For the infinite
plane there is no such length scale whereas in the spherical case this
length scale is the radius of the heat producing sphere. This
advantage is however also the weakness of the method since it limits
the practical frequency window to be studied.

\section*{Acknowledgments} 
The authors thank Prof. Jeppe C. Dyre for supporting this work.  This
work was supported by the Danish National Research Foundation's DNRF
center for viscous liquid dynamics ``Glass and Time''.

\clearpage

\appendix

\begin{figure}
  \centering
  \includegraphics[width=8.6cm]{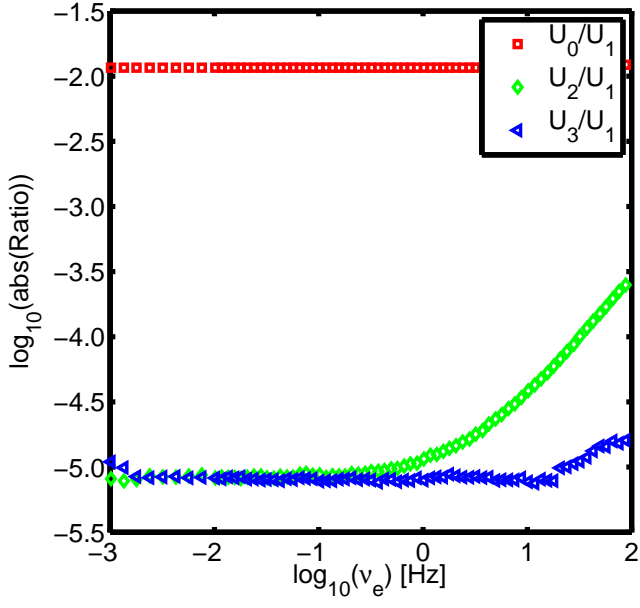}
  \caption{Relative magnitude of the components of input voltage $U$
    to the dominant $U_1$ component. $\langle|U_1|\rangle=4.909\volt$.
    See also table \ref{tab:UFCtab}.}
  \label{fig:UFC}
\end{figure}

\begin{figure}
  \centering
    \includegraphics[width=8.6cm]{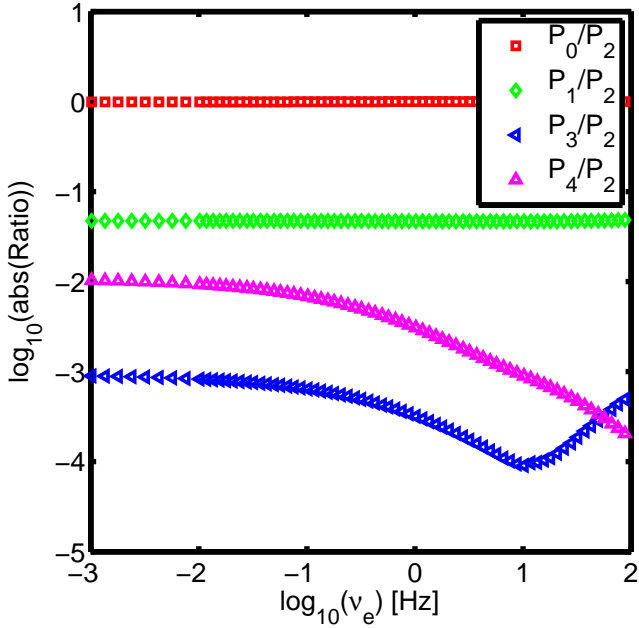}
    \caption{Magnitude of the components of the power to the dominant
      $P_2$ component. $\langle|U_1|\rangle=3.927\volt$,
      $T_\text{cryo}=249.5\kelvin$. See also table \ref{tab:PFCtab}.}
  \label{fig:PFC}
\end{figure}

\begin{widetext}
\section{The AC-method to higher order}
\label{app:ACHigher}
In this appendix we analyze which higher-order terms must be included
in order to get a relative accuracy of $1\E{-4}$ on the measured
temperature amplitude from the Fourier components of the measured
voltages of the voltage divider (see Fig.\ \ref{fig:vdiv}).

In the following we will include up to fourth harmonics in all the
relevant quantities, hence we write
\begin{subequations}
\begin{eqnarray}
  V(t)&=&\frac{1}{2}\left(V_0 +V_1E_1+V_2E_2+V_3E_3 +V_4E_4+c.c. \right)\label{eq:VExpsum}\\
  U(t)&=&\frac{1}{2}\left(U_0 +U_1E_1+U_2E_2+U_3E_3 +U_4E_4+c.c. \right)\label{eq:UExpsum}\\
  P(t)&=&\frac{1}{2}\left(P_0 +P_1E_1+P_2E_2+P_3E_3 +P_4E_4+c.c. \right)\label{eq:PExpsum}\\
  \Delta T(t)&=&\frac{1}{2}\left(T_0 +T_1E_1+T_2E_2+T_3E_3 +T_4E_4+c.c. \right).\label{eq:TExpsum}
\end{eqnarray}
\end{subequations}

Combining the expansion of the thermistor resistance to second order,
Eq.\ (\ref{eq:Rsecond}), with the voltage divider equation, Eq.\
(\ref{eq:VoltageDiv}), the voltage across the preresistor becomes, to
second order in $\Delta T$,
\begin{equation} \label{eq:VSecondOrder}
 V(t)=\frac{1}{A+1}\left(1-a\Delta T+b\Delta T^2\right)U
\end{equation}
with $A=R_0/R_\text{pre}$, $a=\frac{A\alpha_1}{1+A}$ and $b=\left(\frac{A\alpha_1}{1+A}\right)^2-\frac{A\alpha_2}{1+A}$.

If this expression is explicitly calculated using the above expansions
of $U$ and $T$ (Eq.\ \eqref{eq:UExpsum} and \eqref{eq:TExpsum}) a
large number of terms is obtained. In the following we make a
numerical inspection of which terms are most significant, in a worst
case, to reduce the number of terms included.

In order to calculate the order of magnitude of the terms, an estimate
of the size of the components of $U$ and $\Delta T$ are needed.  In
Fig.\ \ref{fig:UFC} we show the relative magnitudes of the different
components of the input voltage, $U$, of the voltage divider to
$U_1$. In table \ref{tab:UFCtab} we summarize the upper limits of
these ratios. In Fig.\ \ref{fig:PFC} we show the relative magnitude
of the different components of the power calculated from the measured
$U_i$ and $V_i$ components. In table \ref{tab:PFCtab} we summarize the
upper limits on these ratios. We estimate the relative size of the
temperature amplitudes as equal to the relative sizes of the power
amplitudes. In table \ref{tab:BeadChar} we give characteristic values
of the quantities describing the thermistor and the resistors.

Combing the estimates given in table \ref{tab:BeadChar}, with Eq.\
(\ref{eq:V3first}), we observe that a change in $T_2$ of $1\E{-4}\kelvin$
corresponds to a change in $V_3/U_1$ of $5\E{-7}$. Hence only terms in
the expansion of Eq.\ (\ref{eq:VSecondOrder}) which are larger than
$5\E{-7}U_1$ will be included.

By explicit substitution of Eq.\ \eqref{eq:UExpsum} and
\eqref{eq:TExpsum} into Eq.\ \eqref{eq:VSecondOrder} and insertion of
the estimates from table \ref{tab:UFCtab}, \ref{tab:PFCtab} and
\ref{tab:BeadChar} the following expressions are found for the first and
third harmonics on the voltage, $V$, when disregarding terms smaller
than $5\E{-7}U_1$

\begin{subequations}\label{eq:V1V3_Full}
\begin{equation}
V_1=\frac{1}{A+1}\left[U_1-aT_0U_1-\frac{1}{2}aT_2U_1^*+X_1\right]
\end{equation}
and
\begin{equation}
V_3=\frac{1}{A+1}\left[-\frac{1}{2}aT_2U_1+U_3+X_3 \right]
\end{equation}
with
\begin{eqnarray}
  X_1&=&-aT_1U_0+bT_0^2U_1+\frac{1}{2}bT_2T_2^*U_1+bT_0T_2U_1^*\\
  X_3&=&-\frac{1}{2}aT_4U_1^*+\frac{1}{4}bT_2^2U_1^*+bT_0T_2U_1+bT_0T_4U_1^*.
\end{eqnarray}
\end{subequations}

\section{Power terms to higher order}
\label{app:power-terms-higher}
The relevant set of power components are calculated from 
Eq.\ (\ref{eq:Palternative}) as 
\begin{eqnarray*}
  P=\frac{(U-V)V}{R_\text{pre}}
\end{eqnarray*}
using the expansions in Eq.\ \eqref{eq:VExpsum} and \eqref{eq:UExpsum}
to be

\begin{subequations}\label{eq:FullPs}
\begin{eqnarray}
  P_0 &=&\frac{1}{R_\text{pre}}\left( W_0V_0+\frac{1}{2}\Re\left(W_1V_1^* +
      W_2V_2^* + W_3V_3^* + W_4V_4^*\right)\right)\\
  P_1 & =&\frac{1}{R_\text{pre}}\left(W_{{1}}V_{{0}}+
    V_{{1}}W_{{0}}+\frac{1}{2}\left(
      W_{{2}}V_{{1}}^*+W_{{1}}^*
      V_{{2}}+W_{{3}}V_{{2}}^*+W_{{2}}^*V_{{3}} + W_3^*V_4+W_4V_3^*\right)\right) \\
  P_2 & =&\frac{1}{R_\text{pre}}\left(
    W_{{0}}V_{{2}}+W_{{2}}V_{{0}}+
    \frac{1}{2}\left(W_{{1}}V_{{1}}+W_{{1}}^*V_{{3}}+W_2^*V_4+W_{{3}}V_{{1}}^*
      +W_4V_2^* \right)\right)\\
  P_3 & =&\frac{1}{R_\text{pre}}\left(
    W_{{0}}V_{{3}}+W_{{3}}V_{{0}}+\frac{1}{2}\left(W_{{1}}V_{{2}}+W_{{2}}V_{{1}}
    +W_1^*V_4 + W_4V_1^*\right)\right)\\
  P_4 & =&\frac{1}{R_\text{pre}}\left(W_0V_4+W_4V_0+\frac{1}{2}\left(
      W_1V_3+W_2V_2+W_3V_1 \right)\right)
\end{eqnarray}
\end{subequations}
where $W=U-V$.

\section{The ``five parameter'' formulation of the thermal transfer
  model}\label{sec:full-five-parameter}
Eq.\ (\ref{eq:TESG2}) , (\ref{eq:Trmth}) and(\ref{eq:Zr0(Zliq)})
are in the following written in terms of the five variables
\begin{eqnarray}
  \label{eq:5parmsApp}
  \tau_l=r_1^2\frac{c_l}{\lambda},\quad
  \tau_b=r_1^2\frac{c_b}{\lambda_b},\quad  \hat{r}=\frac{r_0}{r_1},
  \quad \hat{c}=\frac{c_l}{c_b} ,\quad Z_{\text{liq},0}=\frac{1}{4\pi\lambda r_1}.
\end{eqnarray}

The thermal impedance, Eq.\ (\ref{eq:TESG2}), is given as
\begin{eqnarray}\label{eq:Zliq5parm}
  Z_{\text{liq}}=Z_{\text{liq},0}\frac{1}{\left(1+ \sqrt{i\omega \tau_l}\right)}\,
\end{eqnarray}

The components of the thermal transfer matrix \cite{chr08b} in Eq.\
(\ref{eq:Trmth}) are given as

\begin{subequations}\label{eq:TrmthFull5parm}
\begin{eqnarray}
  T^{\rm th}_{11}
  \,&=&\, \hat{r}\cosh\big(\sqrt{i\omega\tau_b}(1- \hat r)\big) +
  \frac{1}{\sqrt{i\omega\tau_b}}\sinh\big(\sqrt{i\omega\tau_b}(1-\hat
  r)\big)\\
  T^{\rm th}_{12}
  \,&=&\, -Z_{\text{liq},0} \frac{\hat c}{\tau_l}  \frac{\sqrt{i\omega \tau_b}}{\hat
    r}{\sinh\big(\sqrt{i\omega\tau_b}(1-\hat r)\big)}{}\\
   T^{\rm th}_{21}
  \,&=&\, \frac{1}{Z_{\text{liq},0}}\frac{\tau_l}{\hat c} \frac{1}{\left(\sqrt{i \omega \tau_b}\right)^3}
  \Big[\big(1-i\omega\tau_b\hat r\big)
  \sinh\big(\sqrt{i\omega\tau_b}(1-\hat r)\big) -
  \sqrt{i\omega\tau_b}(1-\hat
  r)\cosh\big(\sqrt{i\omega\tau_b}(1-\hat r)\big)\Big]\\
  T^{\rm th}_{22}
  \,&=&\, \frac{1}{\hat r}\cosh\big(\sqrt{i\omega\tau_b}(1-\hat
  r)\big) - \frac{1}{\hat
    r\sqrt{i\omega\tau_b}}\sinh\big(\sqrt{i\omega\tau_b}(1-\hat r)\big)
\end{eqnarray} 
\end{subequations}

\end{widetext}

\clearpage

\begin{table}
  \centering
 \caption{Maximum of relative magnitude of the components of input voltage $U$
    to the dominant first harmonic $U_1$, as shown on Fig.\
    \ref{fig:UFC} (for $\nu<40\hertz$).} 
  \label{tab:UFCtab}
  \setlength{\tabcolsep}{0.7em}
 \begin{tabular}{c c c c}
   \hline \hline 
    $U_0/U_1$ &  $U_2/U_1$ & $U_3/U_1$ & $U_4/U_1$\footnote{$U_4/U_1$
      was estimated from an additional measurement where $U_4$ was
      included.}
 \\ \hline 
    $2\E{-2}$ &  $10^{-4}$ & $10^{-5}$ & $3\E{-5}$ \\
       \hline \hline 
  \end{tabular}
\end{table}

\begin{table}
  \centering
 \caption{Maximum of relative magnitude of the components of the
    power as shown on Fig.\ \ref{fig:PFC}, (for $\nu<40\hertz$).} 
  \label{tab:PFCtab} 
  \setlength{\tabcolsep}{0.7em}
 \begin{tabular}{c c c c}
      \hline \hline 
    $P_0/P_2$ &  $P_1/P_2$ & $P_3/P_2$ & $P_4/P_2$ \\ \hline
    $1$ &   $10^{-1} $ & $10^{-3}$ & $10^{-2}$ \\
     \hline \hline
  \end{tabular}
 \end{table}

\begin{table}
  \centering
 \caption{Characteristic sizes of the parameters describing the used
    thermistor at $T_\text{cryo}=249.5\kelvin$} 
  \label{tab:BeadChar} 
  \setlength{\tabcolsep}{0.7em}
 \begin{tabular}{c c c}
   \hline \hline
   $A$& $\alpha_1$ & $\alpha_2$\\ \hline
   $\approx 2$ & $\approx
   4\E{-2}\reciprocal\kelvin$ & $\approx 10^{-3}\kelvin^{-2}$\\
      \hline \hline
  \end{tabular}
 \end{table}

\clearpage


\end{document}